\documentclass[multphys,vecphys]{svmult}

\usepackage{makeidx}         
\usepackage{graphicx}        
\usepackage{multicol}        
\usepackage[bottom]{footmisc}

\makeindex             

\begin{document}

\title*{Cosmological Interpretation from  High Redshift Clusters Observed Within the XMM-Newton $\Omega$-Project}
\titlerunning{Cosmology with distant XMM-Newton Clusters}
\author{Blanchard Alain\inst{1}}
\institute{$^1$LATT, UPS, CNRS, UMR 5572, 14 Av Ed.Belin, 
 31~400 Toulouse, France
\texttt{alain.blanchard@ast.obs-mip.fr}
}%
%
\maketitle
\begin{abstract}
During the last ten years astrophysical cosmology has brought three remarkable 
results of deep impact for 
fundamental 
physics: the existence of non-baryonic dark matter, the (nearly) flatness of 
space, the domination of the density of the universe by some gravitationally 
repulsive fluid. This last result is probably the most revolutionizing one: 
the scientific review {\it Sciences} has considered twice results on this 
question as {\it  Breakthrough of the Year} (for 1998 and 2003). However, 
direct evidence of dark energy are still rather weak, and the strength of
the standard scenario relies more on the ``concordance'' argument rather than 
on the robustness of direct evidences. Furthermore, a scenario can
be build in an Einstein--de Sitter universe, which reproduces as well as 
the concordance model the following various data relevant to cosmology: 
WMAP results, large scale structure
of the universe, local  abundance of massive clusters, weak lensing 
measurements, most Hubble constant measurements not based on stellar 
indicators. Furthermore, recent data on distant x-ray clusters obtained from 
XMM and Chandra indicates that the observed abundances of clusters
at high redshift taken at face value favors an Einstein de Sitter model and 
are hard to reconcile with the concordance model. It seems wise 
therefore to consider that the actual existence of the dark energy 
is still an open question.
\end{abstract}
\section{Introduction}
\label{sec:1}
\subsection{On the determination of cosmological parameters}
\label{sec:1.1}
The determination of cosmological parameters has always been a central question
in cosmology. However, this problem has become more and more important in 
recent years due to the deep implications it can lead to. One of the most 
spectacular 
results established  in recent years are for instance the existence of a 
dominant form of non-baryonic matter  in the clustered content of the universe.
After a very long debate on whether evidence for non baryonic dark matter 
universe were sufficiently robust, it is nowadays almost unanimously admitted
that there are enough evidences to consider it as an established fact 
(such a conclusion has strongly contributed to emphasize
 the deep couplings that exist 
between astrophysical cosmology and fundamental physics). 
As long as no direct evidence is found (from laboratory experiments) doubts 
are  still possible and indeed few researchers still maintain the point of view 
that modified 
theories of gravitation could do the job as well.\\
  
A second essential result in recent modern cosmology is the evidence 
for the (nearly) flatness of the Universe which 
 comes from the $C_l$ curve of the CMB. The Saskatoon experiment
was probably the first one to provide  evidence for the presence of a peak around 
$l \sim 200$ \cite{net95}, which was shown to provide a 
statistically significant indication for the flatness 
of the universe, a conclusion drawn as early as 1997: 
\cite{lb3} see 
also  \cite{ha98}. This conclusion
 has been   firmly
established by second generation experiments, including  those of 
Boomerang \cite{boom},
Maxima \cite{maxima},  DASI \cite{DASI}, Archeops \cite{archeops}, allowing tight cosmological 
constraints \cite{archeopsPar}. Of course all these results have been 
superseded by WMAP measurements \cite{wmap1,wmap2}. It should be realized
that these CMB measurements provide an observation (basically the position of
the Doppler peak) which is predicted by models, involving standard physics,
consistent with flat models. It is not a direct measurement of curvature
of space (as could be obtained from a triangulation measurement for instance).
The two above results are therefore the unavoidable consequences of the 
existing observations, if they have to be interpreted within standard physics 
as we know it by now. Rejecting these conclusions is possible, but only at the 
expense of modifying fundamental laws of physics as we know them by now. \\

The third result which has emerged in recent years, and which is 
 revolutionizing  for fundamental physics : the dominance of the density of 
the universe by some ``dark energy'', i.e. a fluid with very exotic equation 
of state: $p = w \rho c^2$ with $w \sim -1$ 
\cite{2003A&A...405..409D,2003ApJS..148..175S}. There is a large consensus 
around this so-called concordance model, which 
leads to the idea that the determination of cosmological parameters
has been achieved with a rather good precision, may be of the order of 10\%.
 Indeed  this model does  fit an impressive set of
independent data, the most impressive been: local estimation of the 
density of the universe, CMB $C_l$ curve, most current 
matter density estimations, 
Hubble constant estimation from HST, apparent acceleration of the Universe,
good matching of the power spectrum of matter fluctuations. 
However, the necessary introduction of a non zero
cosmological constant is an extraordinary new mystery for physics,
or more exactly the come back of one of the ghost of modern physics
since its introduction by Einstein. Here the situation is slightly different
from the two previous cases: the introduction of a non-vanishing 
cosmological constant is a major modification of a 
fundamental law of  physics (gravity).
Although the cosmological constant certainly allows one to fit easily 
the Hubble diagram of distant SNIa, its introduction is not unavoidable,
given the data. Rather, trusting fundamental physical laws as we know them 
lead to the conclusion that distant SNIa are, for some unknown reason,
 intrinsically fainter than local ones. There is no argument that dismiss
this ``fact''. Therefore, in order for the SNIa Hubble diagram to be regarded 
as a convincing evidence for a cosmological constant, one should provide a
convincing  independent evidence that the luminosity of a 
distant SNIa is directly comparable to a local one without any correction.
It is therefore the opinion of the author, that in order 
to consider SNIa as an argument
in favor a cosmological constant, evidence for the absence of astrophysical 
corrections to SNIa luminosity has to be demonstrated (and one should remember
 there that the ``absence of evidence `` is not ``an evidence of absence''...).\\

\begin{figure}
\centering
\includegraphics[height=6cm]{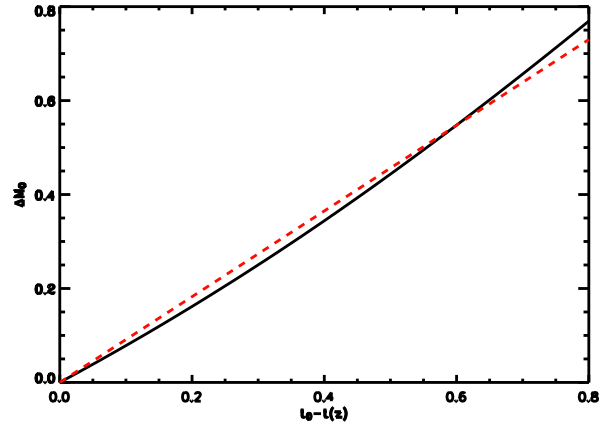}
%
%
\caption{Difference of magnitude between an Einstein de Sitter Universe and 
the concordance (full line) versus look back time. Any process which would 
produce an apparent dimming proportional to this look back time may mimic
the presence of a cosmological constant (dashed line).}\label{fig:sn}
\end{figure}

The possible detection of a cosmological constant from distant supernovae
has brought the first direct piece of evidence largely comforting the
so-called concordance model:  the apparent luminosity of distant supernovae
now appears  fainter, i.e. at larger distance, than expected in any 
decelerating 
universe \cite{riess98,P99} and can therefore  be 
explained
only within an accelerating universe. There is a set of 
 fundamental assumptions in 
this reasoning, that is that SNIa are standard candles which are not affected 
by any bias, any evolution, any obscuration. Although the data are well 
consistent with this hypothesis, it is almost impossible to demonstrate that 
it is actually right, i.e. that data are not biased by some astrophysical 
process. A more problematic point is that astrophysical processes in an 
Einstein de Sitter universe, if roughly
proportional to the look back time may mimic rather well an apparent 
cosmological constant, producing an Hubble diagram that is almost 
indistinguishable from
the standard diagram of the concordance model (see figure\ref{fig:sn}). 
This means that SNIa argument is relatively weak by itself. 
For instance would the SNIa Hubble diagram points toward a negative matter 
content, $\Omega_m < 0.$, it would probably be interpreted by everybody 
as an evidence for some astrophysical process affecting SNIa luminosity...\\ 

I have already discussed in some detail
the various arguments that may raise doubts on the validity of the concordance 
model \cite{Faro02,Porto02} (before WMAP results): for most observations 
which match the concordance model,
there is some other evidences which go in a different direction (for instance
different upper limits on the cosmological constant uncomfortably below 
the present preferred  value were published in the past,
including one coming from the SNIa in the SCP! \cite{P97}.  I would like to add one recent 
example: the Hubble constant. Several measurements based on non stellar 
distance indicators lead to a lower value for Hubble constant than 
that has been derived from HST key-Project \cite{2001ApJ...553...47F}. A recent analysis 
of the Cepheid distances suggests that one bias exist which when corrected 
would lead to a  value 20\% lower for the Hubble constant \cite{2004A&A...413L..31P}. Such a value would imply, in combination with CMB a matter 
density parameter close to $0.5$, ruining the nice concordance of the 
standard paradigm.

\subsection{What the CMB does actually tell us?}
\label{sec:1.2}

\begin{figure}
\centering
\includegraphics[height=8cm]{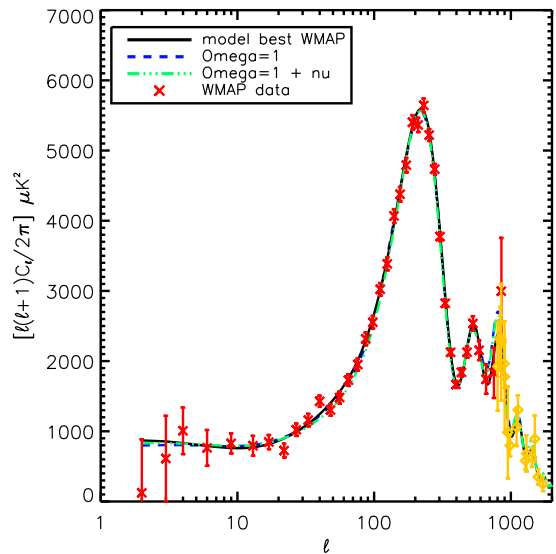}
%
%
\caption{The TT spectrum of the WMAP data compared to three different models: 
one is the concordance, the two others are Einstein de Sitter models, one of 
which  comprises neutrino contribution of $\sim 10\%$ corresponding to three degenerate families with $m_\nu \sim 0.7 $eV. Courtesy of M.Douspis.}\label{fig:dg}
\end{figure}

Since the discovery of the CMB fluctuations by COBE \cite{1992ApJ...396L...1S}
the idea that early universe physics has left imprints revealed by these 
fluctuations has gained an enormous attention. In this respect, 
DMR results have played a fundamental role in modern cosmology comparable
to the discovery of the expansion of the universe or the discovery  of the 
microwave 
background by Penzias and Wilson. The remarkable results of the WMAP 
experiment, are often quoted 
as providing a direct evidence for an accelerating universe. This is 
incorrect: cosmological constraints as established by the WMAP team 
\cite{2003ApJS..148..175S} entirely rely on the powerlaw spectrum asumption.
Therefore these conclusions could be erroneous   
\cite{2001PhRvD..63d3001K,2001PhRvD..63d3009H}. 
Indeed, relaxing this hypothesis, i.e; assuming non power law power spectrum 
allows to 
produce $C_l$ curve which as good as the concordance model. This is illustrated
by figure \ref{fig:dg} on which 3 models are compared to the WMAP data, two 
being 
Einstein de Sitter models. Such models not only reproduce the TT spectrum,
but are also extremely close in term of ET and EE spectra. Furthermore 
the matter power spectrum are similar on scales probed by current galaxies 
surveys. An un-clustered component of matter like a neutrino contribution or
a quintessence field with $w \sim 0$ is necessary to obtain an acceptable
amplitude of matter fluctuations on clusters scales \cite{2003A&A...412...35B}.
Such models require a low Hubble constant $\sim 46$ km/s/Mpc. Such a value
might be look as terribly at odd with central HST key program value ( $\sim 72$ km/s/Mpc) but is actually only $\sim  3 \sigma$ away 
from this value. Given the above mentioned uncertainties (which raised 
the preferred value to lie $\sim  1.75 \sigma$ away, this can 
certainly not be considered as a fatal problem for an Einstein-de Sitter 
universe. The introduction of non-power law power spectrum might appear as 
unnatural. This is a somewhat subjective question. However,  present 
measurement of $C_l$ curve is testing the initial spectrum  over 3 order 
of magnitude in length. The existence of 
distinct features in the primordial spectrum
 are suggested by present WMAP data \cite{Pk}, which could be the consequences 
of early physics on super-Planck scales \cite{Splanck01}, as scales which 
are now accessible to the observations are very likely to be sub-planckian 
before inflation. This argument could be 
regarded as an argument for which non-power-law models are to be preferred 
(although this is not giving 
any support to our --specific-- model, given our poor knowledge of the relevant
 physics). This argument is 
strengthened by the global value of the $\chi^2$ from the WMAP 
$C_l$ : a point that is not much 
emphasized, is that the global value of the $\chi^2$ is not good. In fact, the 
$\chi^2$ for TT data has only a probability of 3\% \cite{2003ApJS..148..175S}. The  conclusion  in such a situation
is that the hypotheses in the  model  are probably to be abandoned! An other 
option is that the data are still suffering from unsubstracted systematics
 (which is the proposed explanation given 
by the WMAP team).

\subsection{Motivation for the XMM-$\Omega$ project}

If one keeps an open mind, one should consider that the existence of a 
cosmological constant is not yet a  scientific fact  established beyond 
reasonable doubt or to be more precise, that the case for its actual  existence
is not as strong as the case for non-baryonic dark matter (furthermore
it is always healthy to have an alternative model to the dominant paradigm).
It is therefore of high interest to have a reliable measurement of 
the matter content of the universe, which in conjunction with the CMB data 
provides a case for or against a non-vanishing cosmological constant, depending
on the value obtained for $\Omega_M$. 
Most of existing measurements are local in 
nature, i.e. they actually provide mass to light ratio (M/L) from  finite and 
relatively small 
entities, 
like clusters, which occupy a tiny fraction of the universe: massive clusters
cover only $10^{-5}$ of the total volume of space! Therefore using 
the  M/L argument relies on an extrapolation over five orders of magnitude...
The baryon fraction has been argued as favoring a low density universe. 
However, this relies on some specific value of the estimation of mass of 
x-ray clusters which is 
uncertain. Consequently, given this uncertainty the 
 baryon fraction is actually consistent as well as 
with a high density universe \cite{sb01}.

The 
evolution of the number of clusters of a given mass is a sensitive function 
of the cosmological density of the Universe, very weakly depending on other 
quantities when properly normalized  \cite{BB}, therefore offering 
 a powerful 
cosmological test  \cite{OB92}.  
The XMM-$\Omega$ project \cite{OP} was designed 
 in order to provide an accurate
estimation of the possible evolution of the luminosity--temperature relation at
high redshift  for clusters of medium luminosity
which constitutes the bulk of X-ray selected samples, in order to remove a
major source of degeneracy in the determination of $\Omega_M$ from
cluster number counts in flux limited number counts.

\section{Observed evolution of the $L-T$ relation of X-ray clusters}

\begin{figure}
\centering
\includegraphics[height=8cm]{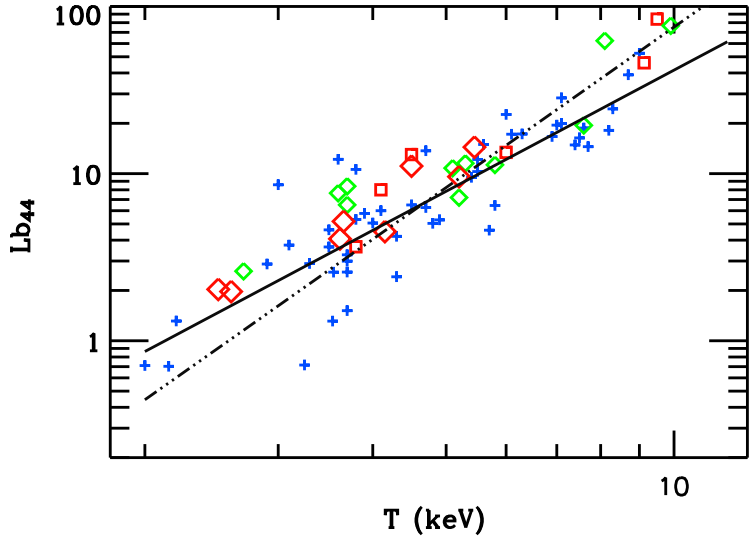}
%
%
\caption{Temperature--luminosity of X-ray clusters:  crosses are local 
clusters from a flux selected sample  \cite{BSBL},  grey diamonds are distant clusters from Chandra  \cite{V02} in the redshift range 
$ 0.4 \leq z \leq 0.625$, large dark diamonds 
are  clusters from the XMM $\Omega$ project, squares are other XMM
clusters within the same  redshift range.}\label{fig:lt}
\end{figure}
For the first time a measurement of the $L-T$ evolution with XMM
has been obtained. D.Lumb et al. (2004)  \cite{OP1} present the results of the X-ray  measurements of 8 distant 
 clusters with redshifts between 0.45 and 0.62. By comparing to various local
$L-T$ relations, clear evidence for evolution
in the $L-T$ relation has been found. The possible evolution has been modeled in the following way:
\begin{equation}
L_{\rm x} = L_{6}(0) \left(\frac{T}{6 \textrm{keV}} \right)^{\alpha} (1+z)^{\beta}
\end{equation}
where $L_{6}(0) \left(\frac{T}{6 keV} \right)^{\alpha}$ is the local $L-T$
relation. $\beta$ is found  to be of the order of $0.6 \pm 0.3$ in 
an Einstein-de Sitter cosmology  \cite{OP1,xmm1}. 
This result is entirely consistent with 
previous analyzes  \cite{sbo,V02} and 
others XMM data (see figure \ref{fig:lt}).
\section{Cosmological interpretation}

\begin{figure}
\centering
\includegraphics[height=4.cm]{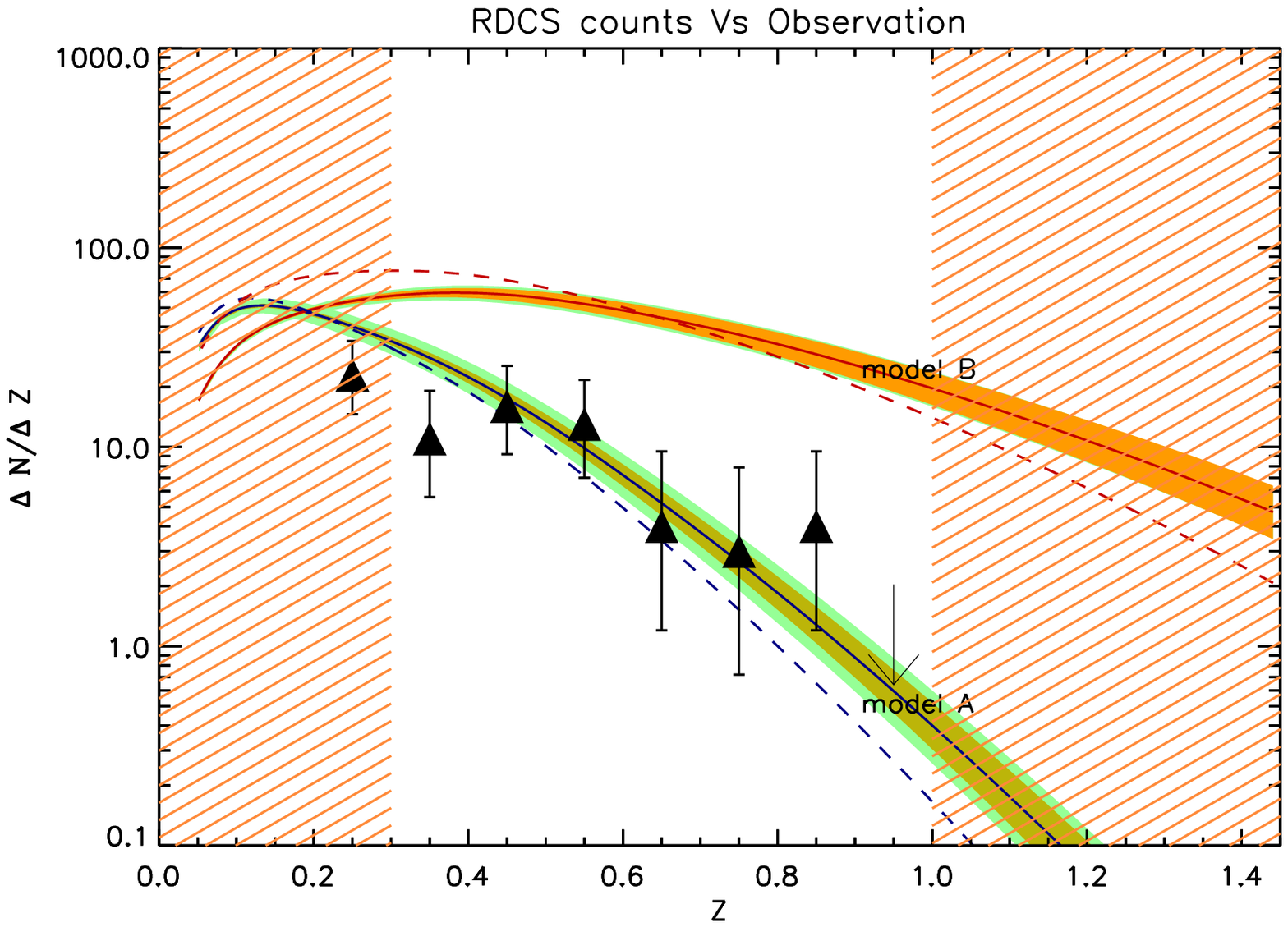}\includegraphics[height=4.cm]{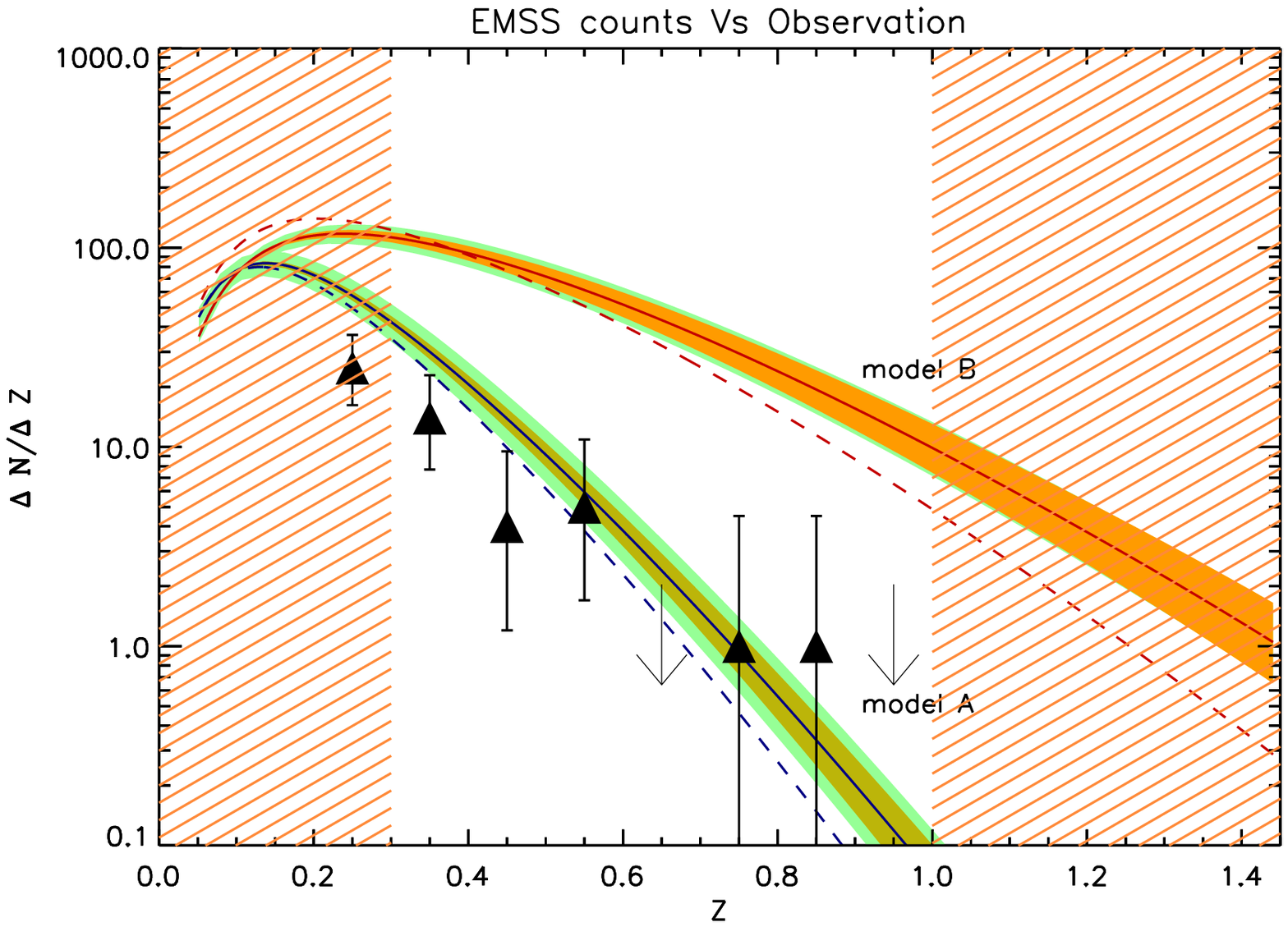}\\
\includegraphics[height=4.cm]{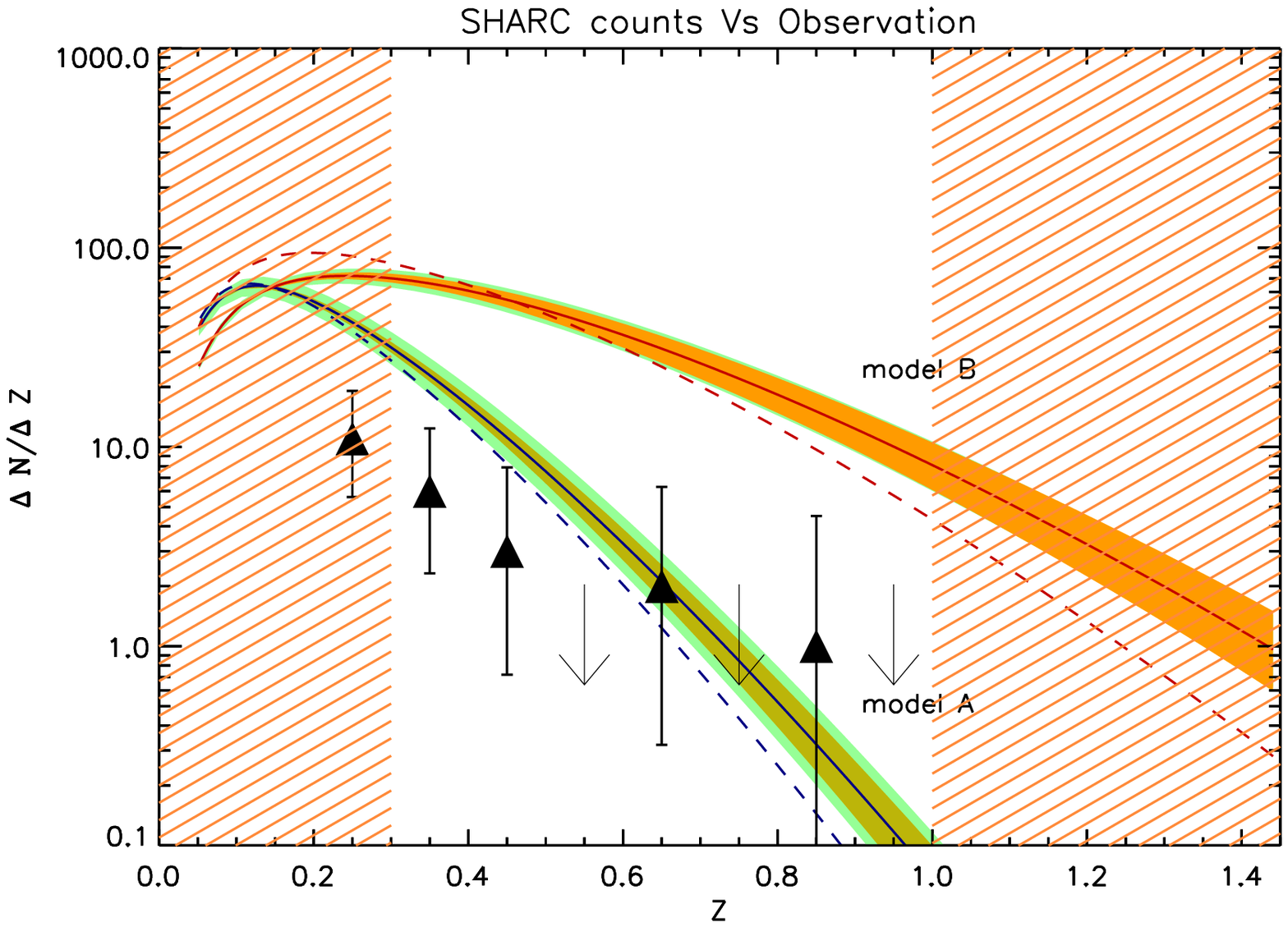}\includegraphics[height=4.cm]{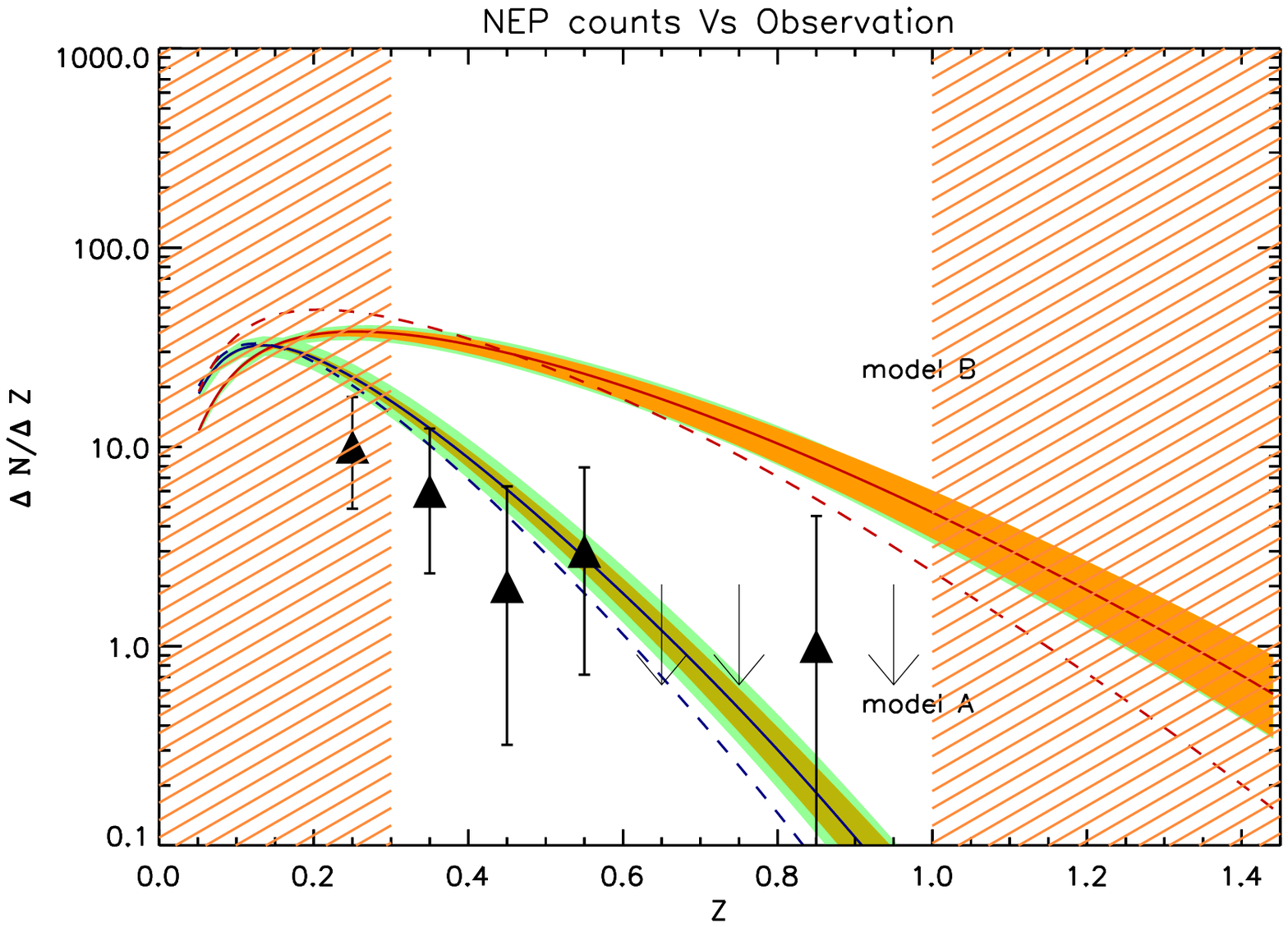}\\
\includegraphics[height=4.cm]{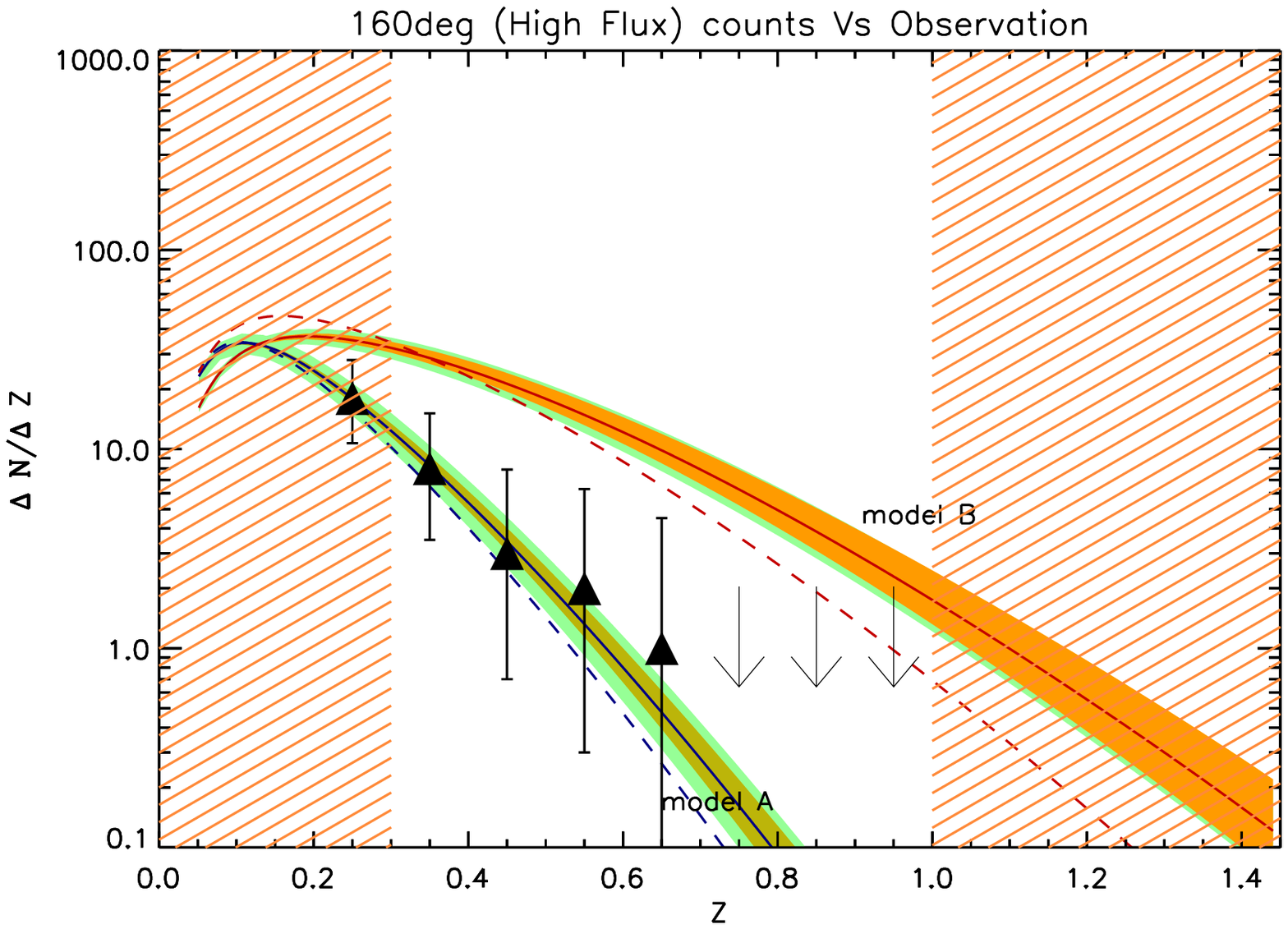}\includegraphics[height=4.cm]{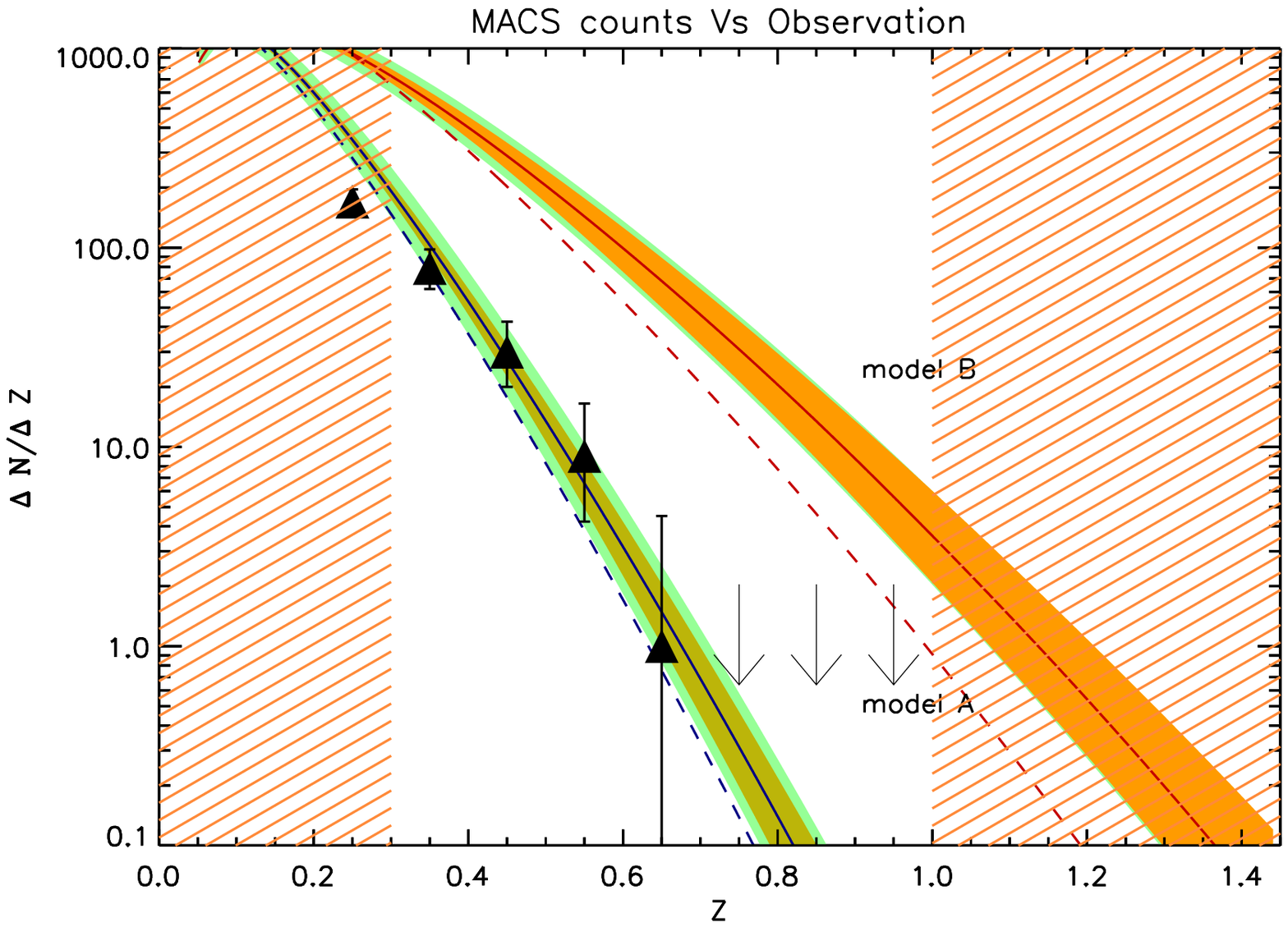}
%
%
\caption{ Theoretical  number counts in bins of redshift ($\Delta z=0.1$) for the  different surveys:
RDCS, EMSS,  MACS and 160deg$^2$-high flux (corresponding to fluxes  $f_x > 2 \, 10^{-13}$ erg/s/cm$^2$).  Observed numbers
are triangles with 95\% confidence
interval on the density assuming poissonian statistics (arrows are 95\% upper limits).
 The
upper  curves are the predictions in the concordance
model (model B). The lower  curves are for critical universe (model A).
Uncertainties on $\sigma_8$ and on $L-T$
evolution lead to the grey area (see \cite{xmm3} ).
}\label{fig:cnts}\end{figure}

\begin{figure}[ht]
\centering
\includegraphics[height=6cm]{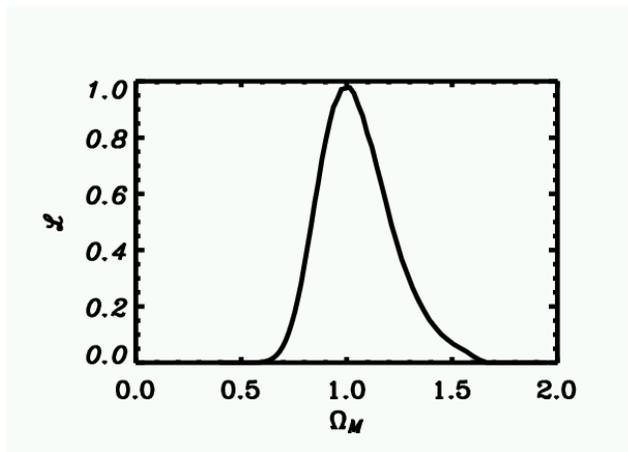}
\caption{ Final likelihood from x-ray cluster number counts obtained with 
independent samples: MACS, EMSS, NEP,  160 deg$^2$. The $M-T$ relation was 
treated self consistently as in \cite{ATM}.
}\label{fig:lik}
\end{figure}

Attempts to apply directly the test of the evolution of the abundance of
clusters have been performed but still from a 
very limited number of clusters (typically 10 at redshift 0.35)  \cite{H97,Eke98,VL99,BSBL}. In  \cite{BSBL} it was found that $\Omega_M = 0.86 \pm 0.25 \ (1 \sigma)$, 
so that a concordance model is away at only a 2-$\sigma$ level, while systematics 
differences explain the values obtained from the various authors. On the 
other hand,
 number counts allow one to use samples comprising much more clusters.
 Indeed using simultaneously different existing surveys: EMSS, SHARC, 
RDCS, MACS
NEP and 160 deg$^2$ \cite{emss92,bsharc,rdcs,macs,nep,vik98} one can use information provided by more than 300 clusters with $z > 0.3$
(not necessarily independent). 
 In order to model clusters number counts, for 
which temperatures are not known, it is necessary to have a good knowledge of
the $L-T$ relation over the redshift range which is investigated, which 
information has been provided by XMM and Chandra. Number counts can then be computed:
\begin{eqnarray}
N(>f_x,z,2\Delta z) = & \Omega \int_{z-\Delta z}^{z+\Delta z} \frac{\partial N}{\partial z}(L_x> 4\pi D_l^2 f_x) dz \nonumber \\
= & \Omega \int_{z-\Delta z}^{z+\Delta z} N(>T(z))dV(z) \nonumber \\
= & \Omega \int_{z-\Delta z}^{z+\Delta z} \int_{M(z)}^{+\infty}
N(M,z)dM dV(z)
\end{eqnarray}
 where $T(z)$ is the temperature threshold corresponding to the flux $f_x$
as given by the observations, being therefore
independent of the cosmological model. For most surveys the above formula has 
to be adapted to the fact that the area varies with the flux limit, and 
eventually with redshift. Several ingredients are needed: the 
local abundance of clusters as given by the temperature distribution function ($N(T)$), the
mass-temperature relation and its evolution, the mass function and  the 
knowledge of the dispersion. Uncertainties in these quantities result in  -systematics-  uncertainties
in the modeling which have been found to be comparable to statistical uncertainties. 
Figure \ref{fig:cnts} illustrates \cite{xmm3} the counts obtained with a standard mass temperature relation:  
\begin{equation}
T =  {\rm 4 keV} M_{15}^{2/3}(1+z)
\label{eq:mt}
\end {equation}
the SMT mass function   \cite{st},
and the $L-T$ relation observed by XMM with its uncertainty. These counts were computed for different existing surveys
to which they can be compared. Several likelihood analyzes have been performed.
Among the various conclusions that were found are: 
all  existing x-ray clusters surveys systematically point toward high 
$\Omega_M$, statistical uncertainties allow a determination of   $\Omega_M$ with a 10\% precision: $ 0.9 < \Omega_M < 1.07 {(1 \sigma)} $. 
During this analysis numerous
 possible source of systematics were investigated with great detail  (local samples, normalization of the 
$M-T$ relation, local $L-T$ relation,  dispersion in the 
various relations).  The dominant source of systematic 
uncertainty is coming from the uncertain calibration of the mass temperature 
relation. This uncertainty can be greatly removed using the method 
based on a self consistent adjustment to the baryon fraction \cite{ATM}. 
With this method the likelihood obtained  is wider and the precision is 
decreased down to 15\% (see figure \ref{fig:lik}).
 In addition the distribution is 
non-Gaussian: with the above prescription, although one conclude that $\Omega_M \sim 0.975 \pm 0.15$, the concordance model
is still ruled out at $7 \sigma$ level. 
  Remaining systematics have been added in quadrature and are 
also representing roughly an additional  15\% uncertainty. This means that 
global uncertainty is roughly 20\%. 
We have also check that the local luminosity in our models
is in good agreement with local surveys (without requesting it explicitly).
 
\section{Looking for loopholes}
\begin{figure}[ht]
\centering
\includegraphics[height=8cm]{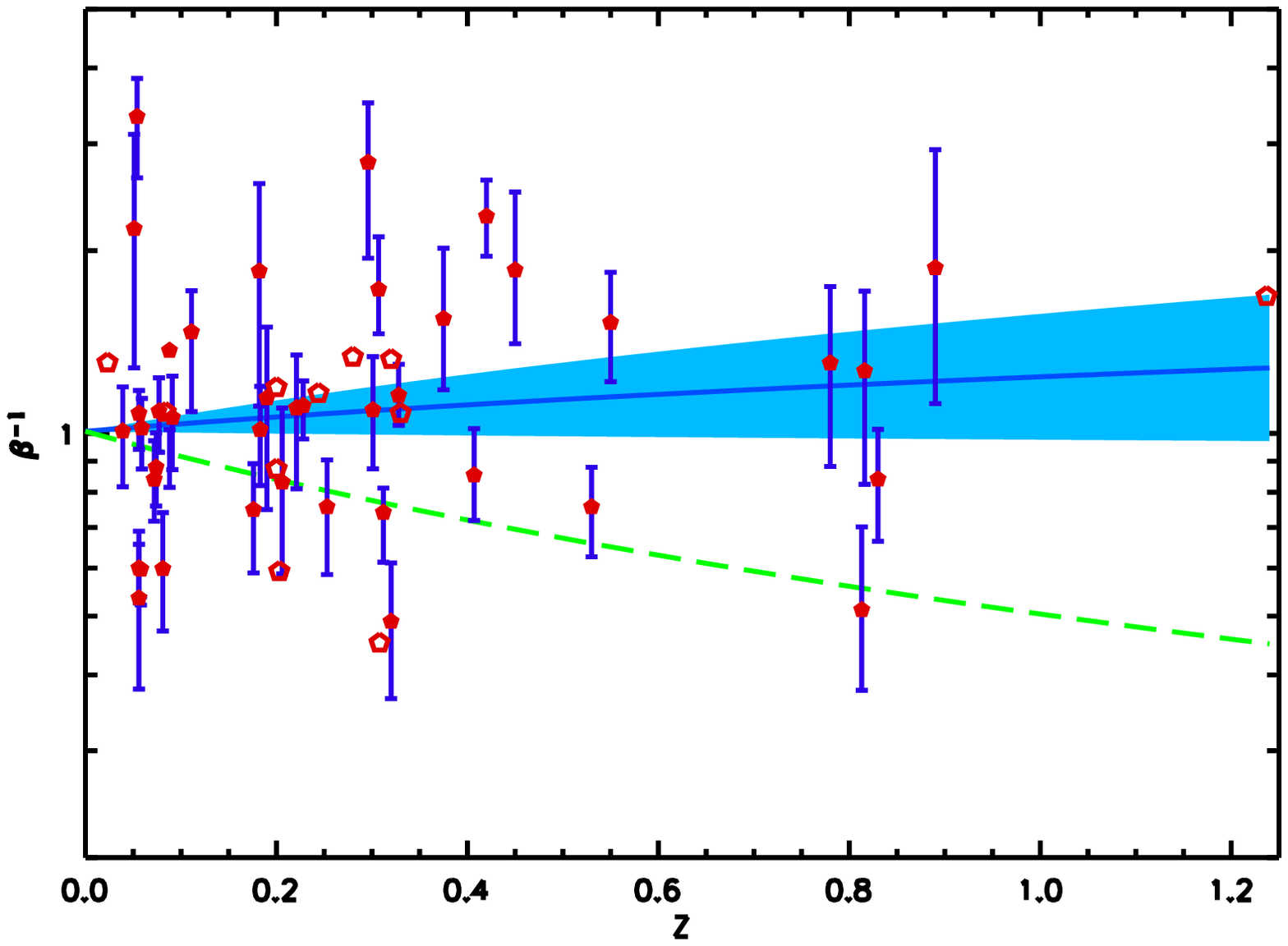}
\caption{The ratio between thermal energy of the gas measured by $T_x$
and the kinetic energy of galaxies measured by their velocity dispersion
for a sample of clusters with $T_x \geq 6$ keV with redshift spanning from 0
to 1.2. No sign of evolution is found. The best fit is the continuous line, 
grey area is the formal one $\sigma$ region, dashed line is the level necessary
to make the concordance in agreement with the x-ray clusters counts.
}\label{fig:vtz}
\end{figure}

\subsection{Systematics}

I have mentioned above that the source of various systematics have been 
investigated and lead to a $\sim 17\%$ uncertainty. This value is 
 larger than the statistical uncertainty $ \sim 10\%$. It is therefore 
very important to investigate one by one this systematics and what 
typical amplitude may restore the concordance. 
 Special attention has been paid to selection functions. For instance if flux 
limit, or identically flux 
calibration in faint surveys, is erroneous by a factor of 2--3 the concordance 
would be much closer to existing surveys. However typical uncertainty is 
considered to be of the order of 20\%. This provides a typical number:
if the value of one of the systematic effects is ten times larger than 
estimated amplitude, then the concordance would accommodate the data. \\

\subsection{Comparison with previous works}

A  comment that is heard sometimes 
in conferences, is that we are the only group
who find such a high value for $\Omega_M$. This an incorrect statement:
when dealing with the $N(T)$ evolution, \cite{VL99} did found a
high central value, close to our best one. Major differences with previous
analyzes to \cite{BSBL} were explained in term of systematics. As those results
lie within the $2 \sigma$ range found in \cite{BSBL}, one can conclude that
the problem is yet open. However, the redshift distribution of X-ray clusters 
using normalization from the local temperature distribution has been 
investigated in the past. With the analysis presented in \cite{OB97,sbo}, 
there 
has been three different independent  analyzes \cite{b99,1999ApJ...518..521R}, each leading to 
consistent 
results with EMSS as well as with ROSAT.  All these analyzes indicate that 
redshift number counts are consistent with a high $\Omega_M$ and at odd with 
value of the order of $0.3$ (note that \cite{b01} have obtained an acceptable 
fit to RDCS distribution, but at the price of unacceptable local abundance).\\

Our new analyzes 
basically recover identical results to the one mentioned above. 
However, the  statistical significance   is now much better: 
these samples contains $\sim$ 300 clusters. Each sample is individually 
well fitted, this is a very important point: 
any large unidentified systematics affecting data, would have to affect 
the different surveys (from different groups and different methodology, on 
both ROSAT and EMSS data) in different way to mimic the Einstein de Sitter 
case, a somewhat tricky coincidence. I conclude that this new analysis is much mores robust than previous one,
both in term of statistic and in term of control on systematics.  

\subsection{Is cluster gas physics essentially non-gravitational?}

We have identified only one possible 
realistic way to reproduce number counts in a concordance model, that is by 
 assuming  
that the redshift evolution of the $M-T$ relation is not standard:
\begin{equation}
T \propto M_{15}^{2/3}
\end {equation}
(i.e. removing the standard $1+z$ factor appearing in equation \ref{eq:mt}). This is conceivably  possible if   
a large fraction of the thermal energy of the gas  in present day clusters 
originates from
other processes than the  gravitational collapse and has been continuously injected during recent past (although it remains to be
shown that this is actually possible in a realistic way).
It is possible to 
test  observationally this latter possibility: heating processes of the gas will 
obviously  heat the gas but not galaxies. The quantity:
$$
\beta^{-1} \propto  \frac{T_x}{\sigma^2}
$$
should therefore evolve with redshift accordingly to $(1+z)^{-1}$ if the $M-T$ relation 
evolved accordingly to the above non-standard scheme while it should  remains 
constant in the standard case. Note that this conclusion persists even if 
galaxies velocity dispersion are a biased version of the dark matter one 
\cite{2004MNRAS.352..535D}.  In order to test whether existing data do provide 
some indication on such a possible evolution,  we have collected 
some existing measurements of velocity dispersion $\sigma$  for massive 
clusters
using BAX cluster data base \cite{BAX} with further recent measurements: we 
selected clusters  with temperature greater than 6 keV for which velocity 
dispersion was available. The result is shown on figure \ref{fig:vtz}. 
 We found no sign of such a  non-standard behavior
which is in principle ruled out at  the 3--$\sigma$ level at least.
\section{Conclusions}
The major results obtained with the $\Omega$ project are the first XMM 
measurement of the evolution of the 
luminosity-temperature with redshift. A positive evolution has been detected, 
in agreement with previous
results including those obtained by Chandra  \cite{V02}. The 
second important result is that this
evolving $L-T$ produced counts in the concordance model which are inconsistent 
with the observed counts in all existing published surveys.
This is in principle the signature of a high density universe, but might be as well due to  a deviation 
of the 
expected scaling of the $M-T$ relation with redshift. Our investigation of the 
ratio $\frac{T_x}{\sigma^2}$  shows no sign of such deviation. Therefore, the 
distribution  of x-ray selected clusters as known at present day favors a 
high density
universe, alleviating the need for a cosmological constant.\\

%
%
%

%
%
%
%
%
%
%

%
%



\printindex
\end{document}